\def\cvprPaperID{0000}
\def\confName{CVPR}
\def\confYear{2023}

\def\paperTitle{Multilayer Perceptron Network Discriminates Larval Zebrafish Genotype using Behaviour}

\def\authorBlock{
    Christopher Fusco\thanks{Equal contribution} \qquad
    Angel Allen\footnotemark[1] \\
    The University of Adelaide \\
    {\tt\small \{christopher.fusco, angel.allen\}@adelaide.edu.au}
}

\newif\ifreview 
\newif\ifarxiv \newcommand{\arxiv}{\arxivtrue}
\newif\ifcamera 
\newif\ifrebuttal 
\arxiv

\pdfoutput=1
\documentclass[10pt,twocolumn,letterpaper]{article}
\ifreview \usepackage[review]{cvpr} \fi
\ifarxiv \usepackage[pagenumbers]{cvpr} \fi
\ifrebuttal \usepackage[rebuttal]{cvpr} \fi
\ifcamera \usepackage{cvpr} \fi

\usepackage{graphicx}
\usepackage{amsmath}
\usepackage{amssymb}
\usepackage{booktabs}
\usepackage{multirow}
\usepackage{graphicx}

\graphicspath{ {./figures/} }

\usepackage{times}
\usepackage{microtype}
\usepackage{epsfig}
\usepackage[table,xcdraw]{xcolor}
\usepackage{caption}
\usepackage{float}
\usepackage{placeins}
\usepackage{color, colortbl}
\usepackage{stfloats}
\usepackage{enumitem}
\usepackage{tabularx}
\usepackage{xstring}
\usepackage{multirow}
\usepackage{xspace}
\usepackage{url}
\usepackage{subcaption}
\usepackage{xcolor}
\usepackage[hang,flushmargin]{footmisc}

\ifcamera \usepackage[accsupp]{axessibility} \fi

% ...

\ifarxiv  \fi

\newcommand{\R}[1]{{%
    \textbf{%
        \ifstrequal{#1}{1}{\textcolor{red}{R#1}}{%
        \ifstrequal{#1}{2}{\textcolor{blue}{R#1}}{%
        \ifstrequal{#1}{3}{\textcolor{magenta}{R#1}}{%
        \ifstrequal{#1}{4}{\textcolor{teal}{R#1}}{%
                           \textcolor{cyan}{R#1}%
        }}}}%
    }%
}}

\usepackage{xr-hyper}

\makeatletter
\newcommand*{\addFileDependency}[1]{
  \typeout{(#1)}
  \@addtofilelist{#1}
  \IfFileExists{#1}{}{\typeout{No file #1.}}
}

\makeatother

\usepackage[pagebackref,breaklinks,colorlinks]{hyperref}
\usepackage[capitalize]{cleveref}
\crefname{section}{Sec.}{Secs.}
\crefname{table}{Table}{Tables}
\crefname{figure}{Fig.}{Figs.}

\begin{document}

\title{\paperTitle}
\author{\authorBlock}
\maketitle

\begin{abstract}

Zebrafish are a common model organism used to identify new disease therapeutics. High-throughput drug screens can be performed on larval zebrafish in multi-well plates by observing changes in behaviour following a treatment. Analysis of this behaviour can be difficult, however, due to the high dimensionality of the data obtained. Statistical analysis of individual statistics (such as the distance travelled) is generally not powerful enough to detect meaningful differences between treatment groups. Here, we propose a method for classifying zebrafish models of Parkinson's disease by genotype at 5 days old. Using a set of 2D behavioural features, we train a multi-layer perceptron neural network. We further show that the use of integrated gradients can give insight into the impact of each behaviour feature on genotype classifications by the model. In this way, we provide a novel pipeline for classifying zebrafish larvae, beginning with feature preparation and ending with an impact analysis of said features.

\end{abstract}


\section{Introduction}
\label{sec:introduction}

Zebrafish are commonly used as model organisms of disease due to their genetic tractability and high reproductive capabilities \cite{newman_using_2014}. To identify new therapeutics for disease treatment and management, high-throughput drug screens can be performed on larval zebrafish in multi-well plates, where some phenotypic response to the treatment is measured. Finding a measurable response can be a challenge, however, which is what we aimed to interrogate here.


The behaviour of zebrafish larvae can be analysed by tracking their movement in a controlled environment and extracting features such as the distance travelled, velocity of movement, or current position in the well. However, analysis of this data is complex as it is the combination of these features that defines behaviour.

The application of machine learning (ML) in biology for classifying model organisms, including zebrafish, is not a new idea in any account. Deep convolutional neural networks trained on coloured images of zebrafish are able to classify them by sex \cite{hosseini_efficient_2019}. The authors did not limit their work to neural networks, however, as they show that sex can also be determined by modelling caudal fin correlation via linear discrimination using support vector machines (SVMs), another well-known supervised learning method. The effectiveness of these approaches are unfortunately unclear due to missing explanation of their method for model training and evaluation.


In this paper, we present a non-linear ML classifier that can predict the genotype of larval zebrafish models of Parkinson's disease (mutation in \textit{dnajc6}, known to cause juvenile-onset PD in humans) with an average of 84\% validation accuracy based on 2D behavioural features. This classifier could potentially be used to indicate whether a mutant larva is displaying a ``healthy" or ``mutant" behavioural phenotype following a certain treatment. We then analyse the impact of behavioural features by calculating the integrated gradients of our models to inform us of the distinguishing features of the genotypes. 

\section{Model Architecture}
\label{sec:model_architecture}
\begin{figure}[t]
    \begin{center}
        \includegraphics[width=1\linewidth]{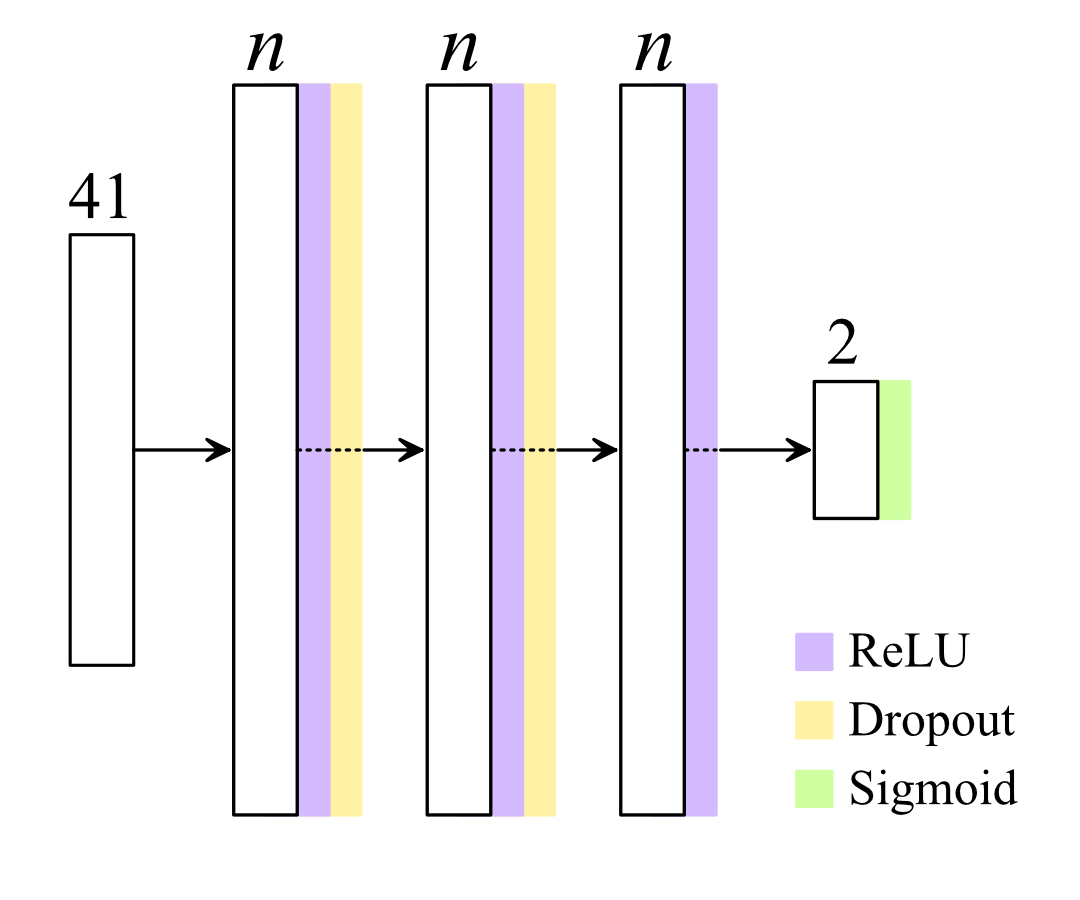}
    \end{center}
        \caption{A visualisation of the network architecture used. The input layer has a perceptron for each behaviour feature (41). Each hidden layer is followed by a ReLU activation function. The first two hidden layers are inhibited by dropout. The output layer has two perceptrons, one for each genotype (wildtype and homozygous), and is passed to a sigmoid activation function.}
    \label{fig:architecture}
    \end{figure}

Our proposed network architecture has an input layer of 41 dimensions, three hidden layers of $n$ dimensions, and an output layer of two dimensions. We chose an output size of two, even though we are solving a binary classification problem, to allow for the simple addition of another genotype for classification (heterozygous). The optimal dimension of the hidden layers is decided after evaluating our experiments as discussed in the next section. The architecture is visualised in Figure \ref{fig:architecture}.

\subsection{Initialisation}
Random initialisation of a neural network has been shown to lead to slower convergence and toward poorer local minima \cite{glorot_understanding_2010}. Xavier initialisation is a popular arrow in the quiver of deep neural network (NN) researchers as it decreases the probability of encountering the vanishing/exploding gradients problem during training \cite{boulila_weight_2021}. However, this method is based on the assumption that are linear \cite{kumar_weight_2017}. This assumption is invalid for the ReLU activation function used in our proposed architecture. He \textit{et al}. proposed a new initialisation strategy for ReLU-activated networks \cite{he_delving_2015}. \textit{He initialisation} sets the network weights randomly from a normal distribution, as in Xavier initialisation, but with a variance of: \begin{equation} v^{2} = 2/N \end{equation} in order to reduce the probability of the dying neuron problem.

\subsection{Generalisation}
Batch size is an important contributor to the generalisation ability of a network. When employing stochastic gradient descent as a learning strategy, it is necessary to ensure that the batch size is not too large so as to not reduce the gradient noise \cite{kandel_effect_2020, he_control_2019}. Fluctuation in gradient is sometimes required to escape a poor local minima, thus reduction of this ability can be detrimental. A batch size of $10$ proved helpful for our network to reach good minima in an adequate amount of epochs.

A significant contribution to the field in terms of improving the generalisation of a network is Dropout \cite{srivastava_dropout_nodate}. The proposed idea is to randomly drop perceptrons and their connections from the network during training. Although simple it is extremely effective and has seen massive adoption to network architectures \cite{labach_survey_2019}. By randomly inhibiting the network it prevents units from co-adapting too much. Metaphorically speaking, it prevents the network from memorising the answers to the test since it is easier than studying the concepts \cite{ying_overview_2019}. We found that inciting a dropout ($p=0.5$) after the first and second layers of our network was ideal for achieving greater generalisation.

\subsection{Integrated Gradients}
Understanding the flow of information through NNs is a challenging problem that has recently gained increased attention. Being able to analyse the impact of input features on outputs allow further analysis and selection of features for obtaining better results \cite{ancona_towards_2018}. A simple manner of evaluating the importance of features given by a network can be by analysing the input gradients, that is, the gradients between the input layer and the first hidden layer. But this has a clear limitation of only representing a small portion of the network, akin to looking at it through a peep hole. 

We make use of a method that satisfies two fundamental axoims of attribution - \textit{Sensitivity} and \textit{Implementation Invariance}. In simpler words, a lack of sensitivity causes gradients to focus on irrelevant features and networks that produce the same outputs, given the same input data, should always have the same attributions. Sundararajan et al. show that most methods do not satisfy these axioms and proposed a new method \cite{sundararajan_axiomatic_2017}. Integrated gradients postulates an additional axiom called \textit{completeness}, stating that, for an input $x$, the attributions should add up to the difference between the output of a network and a baseline. It is suggested that the baseline be chosen such that its prediction is near zero. The proposed technique can be applied to a range of network architectures, including ours.

\subsection{Hyper-Parameters}
We designed a lightweight neural network for classifying wildtype and mutant genotypes. An important design choice was for the network to be able to be trained using a portable laptop, in our case, an M1 Macbook Pro. This was accomplished by only having a hidden depth of three layers that are fully connected. A sufficiently wide neural network with just a single hidden layer can approximate any (reasonable) function given sufficient training data, although increasing the hypothesis space requires an exponential increase in width \cite{eldan_power_2016}. Our hypothesis space is sufficiently small where three layers is adequate to approximate our ideal function.

The output of the final layer in the NN is passes through a sigmoid function to produce a probability distribution. A logistic loss (binary cross entropy) with one-hot encoded labels is used to calculate the error between the predicted distribution and the true distribution. The loss is backpropagated via stochastic gradient descent (SGD) \cite{robbins_stochastic_1951} where, paired with the non-linear activation function ReLU, we are able to reliably find the global minima of the training loss \cite{zou_stochastic_2018}. Adding momentum to SGD has also been shown to be helpful in traversing sub-optimal local minima \cite{sutskever_importance_2013}. A learning rate of $5e^{-4}$ was chosen in combination with a momentum of $0.99$ and proved to be near optimal for our network training.

\section{Methods}
\label{sec:methods}

\subsection{Data Collection}
Zebrafish larvae with wildtype and mutant \textit{dnajc6} genotypes were generated by heterozygous incross of two parent pairs. Two families were generated on subsequent days. Larvae were placed in 24-well plates in 1 mL of E3 embryo medium \cite{westerfield_zebrafish_2000} at 4 days old, before being subjected to behavioural testing at 5 days old. Video footage was collected of larvae swimming in the 24-well plates for 15 minutes using the DanioVision (Noldus). A sudden light to dark transition was used to provoke a stress response in the larvae once 5 minutes had elapsed, as we believed that this stimulation would emphasise any genotypic effects on behaviour (specifically locomotor deficits). Each tray of larvae was subjected to 4 separate behaviour trials throughout the course of the day to augment the data available for the network. Fish were genotyped after behaviour testing.

Tracking of the zebrafish and calculation of behavioural variables was done using the EthoVisionXT software (RRID: SCR\_000441). The ``centrepoint" (trunk) of each larva was tracked by the software from live 2D video footage at 15 frames/second. Table 1 shows the statistics calculated for each 1 minute time bin. Note that there is no correspondence across one minute windows.

\newcolumntype{P}[1]{>{\arraybackslash}p{#1}}
\begin{table}
    \caption{Behavioural variables used to train the network. Features were calculated based on the tracked centre-point of each larva for one minute time bins. For each variable, one or more statistics were calculated for a total of 25 variables.}
    \small
    \begin{tabular}{P{83pt} P{35pt} P{83pt}}
        \toprule
        \textbf{Variable}&\textbf{Units}&\textbf{Statistic}\\
        \midrule
        Distance travelled  & mm        & Total, mean, variance   \\
        Velocity	        & mm/s	    & Mean, variance          \\
        Time moving	        & s	        & Total                   \\
        Frequency moving	& count	    & Total                   \\
        Acceleration	    & mm/s$^2$  & Max, min, variance      \\
        Frequency in middle & count     & Total                   \\
        Time in middle	    & s	        & Total, mean, variance   \\
        Distance to middle	& mm	    & Total, mean, variance   \\
        Mobility	        & \%	    & Total, mean, variance   \\
        Meander	            & deg/mm	& Total, mean, variance   \\
        Heading	            & deg	    & Mean, variance          \\
        \bottomrule
    \end{tabular}
\label{tab:features}
\end{table}

\subsection{Pre-processing}
Pre-processing of the raw data was performed in R. Behavioural data from both families (spawned and tested on different days) was joined to the fish metadata (genotype, tray number, position in tray, trial number). Two fish were omitted, one due to a deformity that prevented movement and the other due to issues with detection of the fish by the behaviour tracking software.

\subsection*{Outlier removal}
The absolute z-score (distance from the mean in standard deviations) was calculated for each data point based on the mean value of that statistic within a genotype group. This was done to prevent the removal of attributes that were due to differences between genotypes. Any values with a z-score $>$4 (4 standard deviations from the mean) were removed. This was done separately for each family before merging all data.

\subsection*{Data summarisation}
The values of each behavioural statistic were averaged for each fish across three time bins from the first light period (bins 3, 4, and 5), and the dark period (bins 6, 7, and 8). These summarised values were used as behavioural features (eg. distance travelled in the light, distance travelled in the dark). A total of 41 features were used here. For the dark period, all statistics in Table 1 were used and for the light period, all statistics other than the distance to middle (variance), meander (all), and heading (all).

\subsection{Model Training}

Prior to training, the data was standardised to have a mean of zero and a standard deviation of one to encourage faster convergence. A random 85/15 train/validation split was done to set aside data for testing the generalisation ability of the models. Each model was trained on the validation data for 500 epochs and the model was evaluated after each epoch. The model with the highest validation accuracy during this time was chosen as the most fit model and was used for calculating integrated gradients.

\section{Results}
\label{sec:results}
\begin{figure}[t]
    \begin{center}
        \includegraphics[width=1\linewidth]{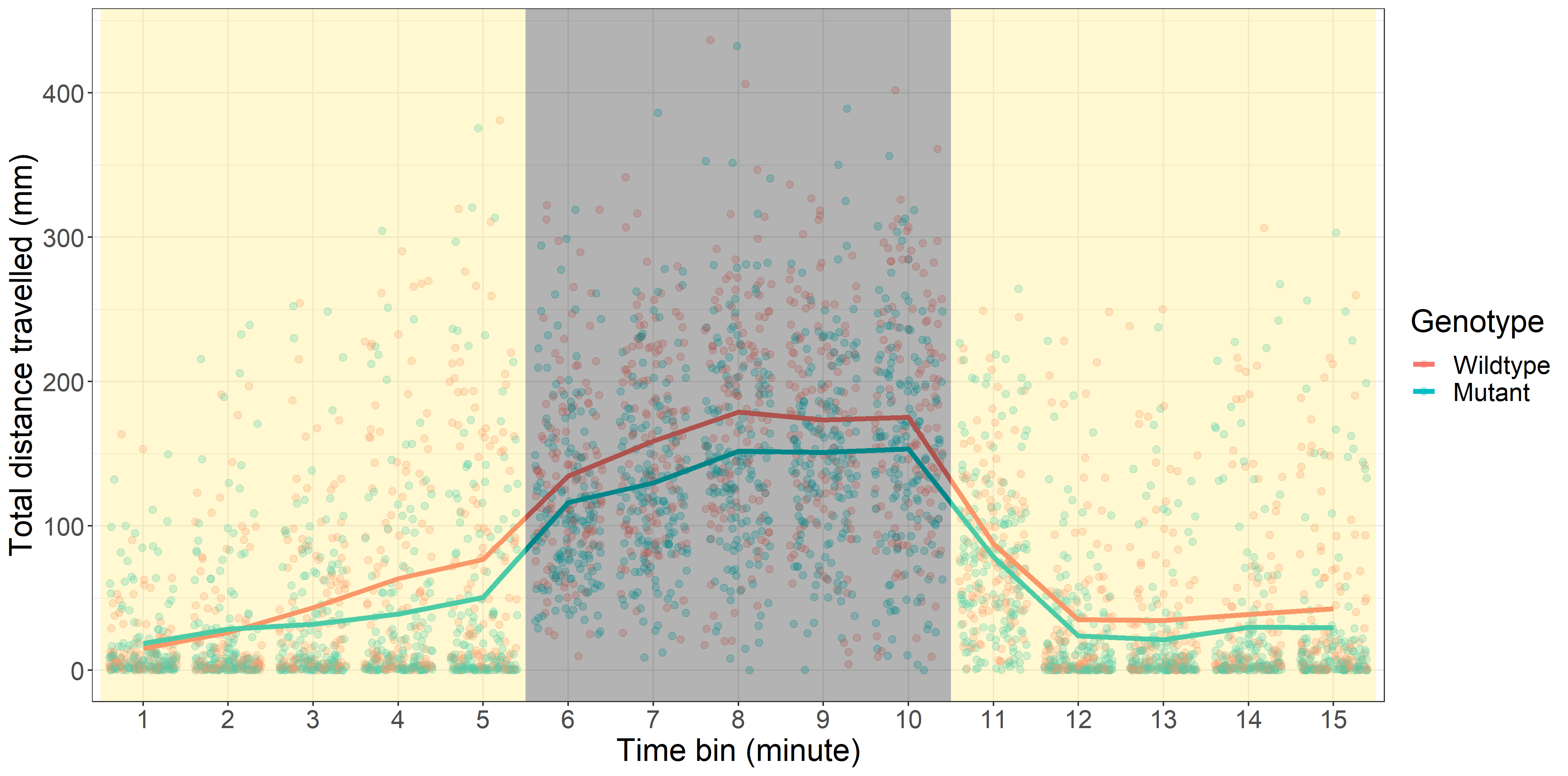}
    \end{center}
        \caption{Total distance travelled (in mm) by larvae for each one minute time bin (total 15 minutes). The coloured panels indicate the light state, with a sudden light to dark transition at the 5 minute mark. Lines indicate the mean for each genotype over time.}
    \label{fig:raw_data}
    \end{figure}

Larvae showed an overall increase in movement following the light to dark transition, shown in Figure \ref{fig:raw_data}. There was a visible difference in the group mean for the total distance travelled during both the light and dark periods between wildtype and mutant larvae, indicating possible locomotor deficits in the mutants. 

During pre-processing of the data, $673$ values were considered outliers and omitted. The majority 

\subsection{Training and Inference}
Ten experiments were run with differing numbers of perceptrons in the hidden layers to allow comparison in terms of overfitting and selection of the best model. We performed each experiment on the same $85/15$ train/validation split so that results were easier to compare. Each experiment trained ten models of the same architecture and were evaluated on the validation set. Early stopping was used to select the best version of each model. 

The average train accuracy and validation accuracy of each experiment for analysis, shown in Table \ref{tab:results}. The architecture with 70 perceptrons in the hidden layers produced the best train accuracy. The 80-perceptron architecture had the best accuracy on the validation set and came a close second on the train set. We found that architectures with more than 30 perceptrons in their hidden layers were more prone to loss divergence but still achieved high train and validation accuracies. In comparison, the architectures with less than 30 perceptrons had a lower frequency of divergence but could not achieve the same accuracy. We theorise that this is due to the networks not having a sufficient number of parameters to fit the data well. In comparison, the losses of  architectures with more perceptrons were more likely to diverge, but were also more likely to achieve better generalisation.


\begin{table}[h]\centering
    \begin{tabular}{ccccc}
        \toprule
        \multirow{2}[3]{*}{\textbf{Perceptrons}} & \multicolumn{2}{c}{\textbf{Train (\%)}} & \multicolumn{2}{c}{\textbf{Validation (\%)}} \\
        \cmidrule(lr){2-3} \cmidrule(lr){4-5}
        & Mean & Stdev & Mean & Stdev \\
        \midrule
        10  & 84.0 & 4.40 & 81.1 & 2.96 \\
        20  & 81.8 & 4.55 & 80.0 & 4.39 \\
        30  & 85.3 & 2.50 & 82.7 & 3.42 \\
        40  & 86.8 & 3.28 & 83.6 & 1.44 \\
        50  & 87.2 & 2.86 & 83.0 & 1.93 \\
        60  & 87.0 & 3.03 & 83.2 & 2.44 \\
        70  & 88.4 & 2.29 & 83.0 & 2.45 \\
        80  & 88.4 & 2.00 & 84.1 & 1.52 \\
        90  & 87.2 & 2.84 & 82.7 & 2.20 \\
        100 & 87.7 & 2.12 & 83.4 & 1.53 \\
        \bottomrule
    \end{tabular}
    \caption{Train and validation accuracies (\%) for ten different experiments with increasing numbers of perceptrons in their hidden layers. Each experiment trained ten models, for which the mean and standard deviation of the train and validation accuracies were calculated.}
    \label{tab:results}
\end{table}

\subsection{Integrated Gradients}
\begin{figure}[ht]
\begin{center}
    \includegraphics[width=1\linewidth]{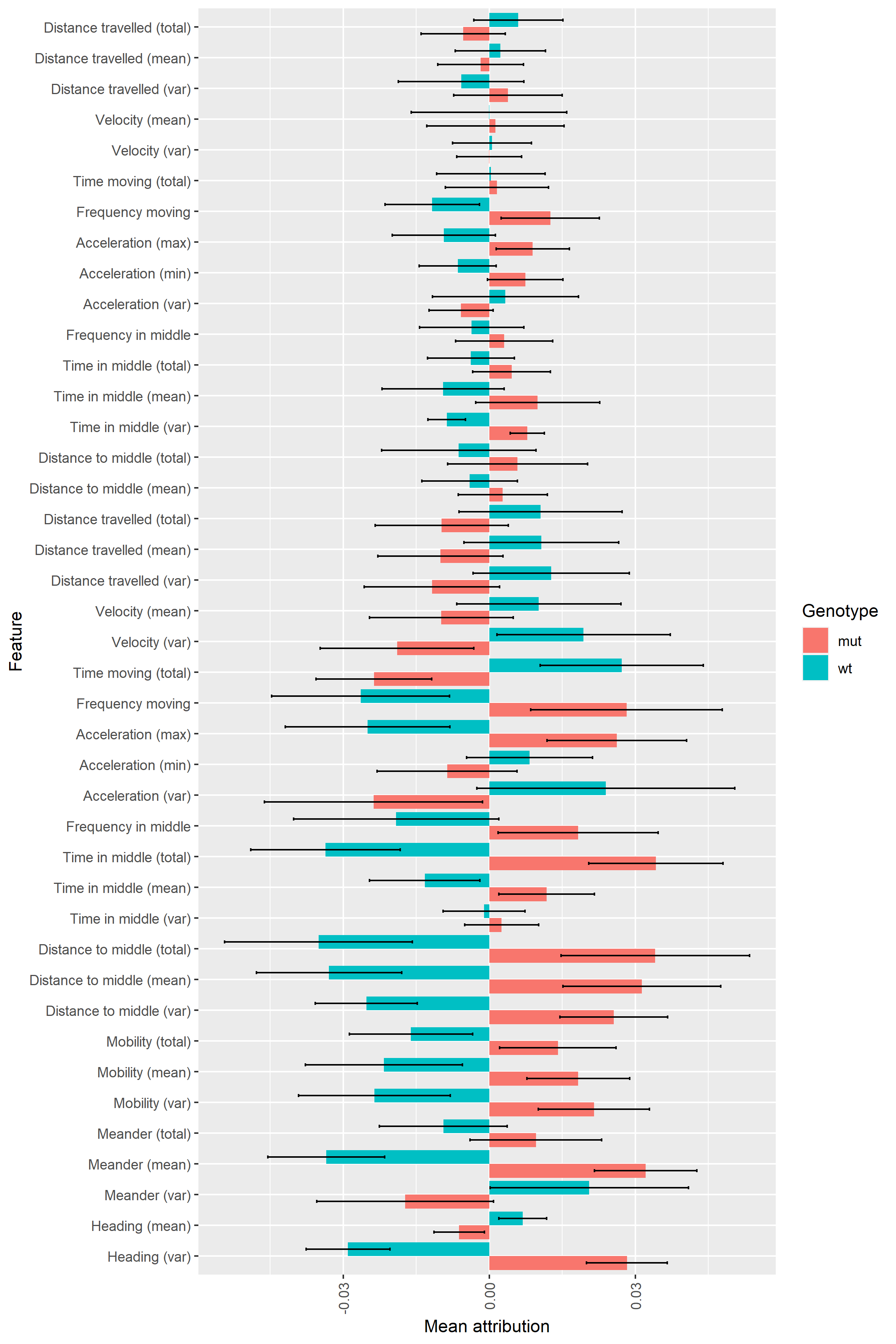}
\end{center}
    \caption{Integrated gradients averaged over ten models with 80-perceptron hidden layers. Large attribution scores (in either direction) indicate a large influence of a feature on the output. Positive/negative attributions indicate that a feature increases/decreases the probability of a certain genotype being chosen. Standard deviation shown as error bars.}
\label{fig:gradients}
\end{figure}

The average integrated gradients from the ten models were calculated for each experiment. Each behavioural feature was given an attribution value representative of the impact it had on the classification outcome across the entire dataset. Positive attributions indicate that the feature increases the probability of the genotype being chosen, and vice versa. The average integrated gradients for all models with hidden layers of 80 perceptrons are visualised in Figure~\ref{fig:gradients}.

The mean distance travelled, mean velocity, velocity variance, time in middle variance, and total time moving have very small attribution scores indicating that the model is generally unsure how to make use of these features. Whereas we can see that the other features have larger attribution scores, indicating that the model depends on them for the label decision.

Overall, the behavioural features from the dark period (after a stressful stimulus) show higher attribution scores than those from the light period. This may be due to the overall increased movement seen in larvae following the light to dark transition (Figure~\ref{fig:raw_data}) providing more information with regards to their movement.

The velocity and time moving in the light have very small attribution scores indicating that the model is generally unsure how to make use of these features. This is also seen in the variance for the time spend in the middle in the dark. Many other features have larger attribution scores, indicating that the model depends on them for the label decision.

\section{Discussion}
\label{sec:discusson}

\begin{figure}[t]
    \begin{center}
    \subcaption[]{\includegraphics[width=1\linewidth]{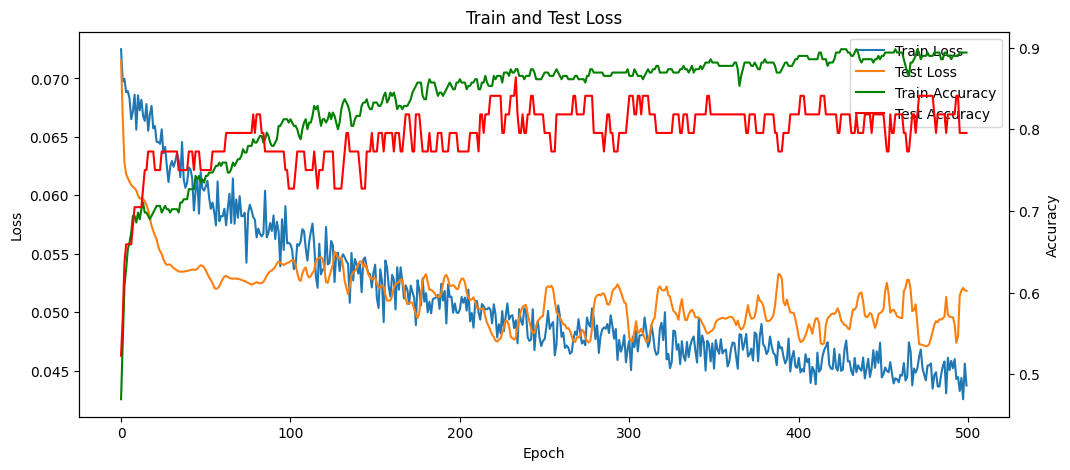}\label{fig:first_sub}}
    \subcaption[]{\includegraphics[width=1\linewidth]{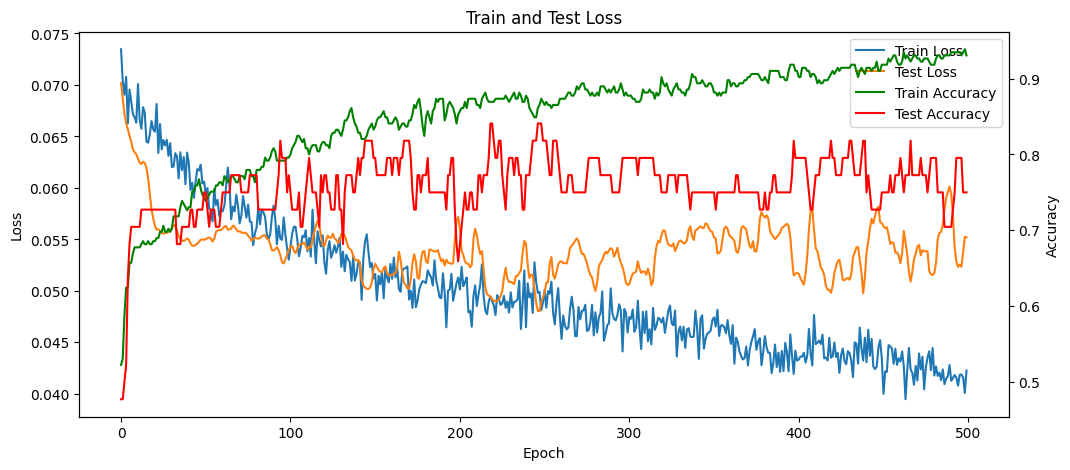}\label{fig:second_sub}}
    \end{center}
        \caption{Visualisation of the train \& validation loss (blue \& orange) with the train \& validation accuracy (green \& red) of a model containing 80 perceptrons in each hidden layer. Both graphs show relatively steady trends for train accuracy and train loss. It is apparent that \ref{fig:first_sub} achieved good generalisation and that there is good convergence of the losses over 500 epochs, although slight divergence is starting to be seen toward the end. \ref{fig:second_sub} has same architecture but is a different instance, showing earlier convergence and again achieving over 80\% validation accuracy, however clear overfitting can be seen starting at epoch 250.}
\label{fig:losses}
\end{figure}

We observed that having a small number (ten) of perceptrons in the hidden layers reduced the loss gap at the cost of overall learning capability. Through our experiments we found that 80 perceptrons in each layer produced the best accuracy on the validation set (Table \ref{tab:results}). The training of this model and another 80-perceptron model is shown in Figure \ref{fig:losses}. The potential for higher accuracy with the drawback of faster overfitting was observed in the models with a larger number of perceptrons, as expected. The inverse was seen in the models with less perceptrons.

The average integrated gradients of each experiment were very similar between models with different hidden layer sizes. Further experimentation with different data splitting techniques, preferably k-fold cross validation, would provide insight to whether these attributions are reflective of the behaviours in a general sense or whether they are different each time the training set changes. Furthermore, investigation into whether consistent attributions correlate with validation accuracy would provide meaningful insight into the usefulness of this technique overall.

\section{Conclusion}
\label{sec:conclusion}

In this work, we show that the classification of zebrafish by genotype using a shallow neural network trained on two-dimensional behaviour features is feasible. The proposed model architecture shows an average 84\% accuracy. Correct initialisation of the network, paired with methods for combatting overfitting such as dropout, tuning the batch size, and optimisation of learning parameters, allowed for successful training of these models.

The integrated gradients of these models provide a way of analysing the impact of behavioural features with respect to its predictive output. The attribution scores can be used to remove low-scoring features that are more likely to confuse the model during training rather than assist with correct label predictions.

\section{Acknowledgements}
\label{sec:thanks}

The authors would like to thank Jack Valmadre and Karissa Barthelson for their guidance in preparing this manuscript.

{\small
\bibliographystyle{ieee_fullname}
\bibliography{references}
}


\end{document}